\documentclass[a4paper]{article}

\usepackage[english]{babel}
\usepackage[utf8x]{inputenc}
\usepackage[T1]{fontenc}

\usepackage[a4paper,top=3cm,bottom=2cm,left=3cm,right=3cm,marginparwidth=1.75cm]{geometry}

\usepackage{amsmath}
\usepackage{graphicx}
\usepackage[colorinlistoftodos]{todonotes}
\usepackage[colorlinks=true, allcolors=blue]{hyperref}

\title{Security Analysis of Near-Field Communication (NFC) Payments}
\author{Dennis Giese, Kevin Liu, Michael Sun, 
Tahin Syed, Linda Zhang}

\begin{document}
\date{May 16, 2018}
\maketitle

\begin{abstract}
Near-Field Communication (NFC) is a modern technology for short range communication with a variety of applications ranging from physical access control to contactless payments. These applications are often heralded as being more secure, as they require close physical proximity and do not involve Wi-Fi or mobile networks. However, these systems are still vulnerable to security attacks at the time of transaction, as they require little to no additional authentication from the user's end. In this paper, we propose a method to attack mobile-based NFC payment methods and make payments at locations far away from where the attack occurs. We evaluate our methods on our personal Apple and Google Pay accounts and demonstrate two successful attacks on these NFC payment systems.
\end{abstract}

\section{Introduction}
Prior to the digital age, physical access control was managed by locks and keys and payments were only made via cash. Today, these are being phased out in favor of digital solutions. Physical access control is now often managed by magnetic, wireless ID cards or biometrics, such as fingerprints, while payments can be made by credit cards or contactless payment methods. The convenience of these new digital methods are making them increasingly popular. However, the greater prevalence of these technologies poses new security risks. Whereas in the past a key or cash might have to be stolen physically, they can now be stolen digitally. Furthermore, rather than just having a key or a few bills stolen, these virtual attacks can wreak havoc on a person's life, leading to a severe invasion of privacy and identity theft. Therefore, with over \$190 billion of transactions by over 60 million users expected to be made by 2020 using just smartphone-based NFC, it is imperative that the security of these technologies be understood by society \cite{nfc-stats}. In this paper, we look specifically at NFC payment methods, focusing our analysis on the security at the point of sale.

In this remainder of this paper, we will present a thorough security analysis of NFC contactless payment methods. Section 2 will detail a survey of previous payment technologies and the security vulnerabilities present in each of these methods. In Sections 3 and 4, we will present an overview of NFC technology from both a general perspective and a smartphone payment-specific security perspective. We will also use these security evaluations to propose potential areas of attack in section 5. Section 6 will present our main results experimenting and evaluating the efficacy of wormhole attacks on NFC payment methods. Using these results, we propose some security recommendations in Section 7 designed to defend against our NFC-based attacks, before some concluding remarks.

\section{Baseline Payment Technologies}

To establish a baseline, we begin with an overview of past technologies used for payment, identification, and access control and the security properties of each.

\subsection{Magnetic Stripe Cards}

Magnetic stripe cards (magstripes) were developed several decades ago \cite{magstripewiki} and have been the dominant technology used for transactions, identification, and access control since. Data on the card is organized in three tracks stored on the stripe, each with specific characteristics in terms of the information density and content it can store. The tracks encode this data by modifying the magnetism of iron-based particles \cite{magstripestandard}. However, the data is static and cannot be dynamically altered to respond to queries from a terminal (ex. even in cases of unauthorized access). As a result, magstripe cards have security analogous to handwritten notes on a sheet of paper. The static data can be written to and read from by anybody with an inexpensive (\$20) skimmer. Furthermore, an adversary can write this data to a blank card and create an effective clone for slightly more (\$60). \cite{magstripewiki}. 

\begin{figure}[h]

\centering
\includegraphics[width=0.6\textwidth]{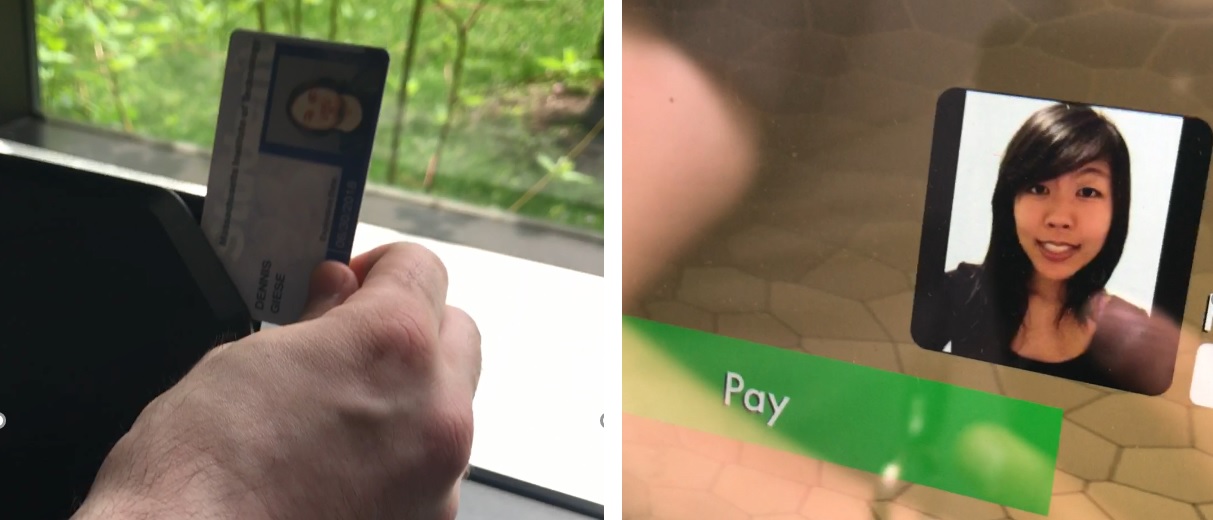}
\caption{\textit{Left}: Dennis swiping his MIT ID card with modified track information at a card reader. \textit{Right}: Linda's image being displayed after Dennis's swiped card.}
\label{fig:idcard}
\end{figure}

As a proof-of-concept, we obtained permission from MIT Card Services to try some exploits on MIT ID cards. We replaced the track information on the magnetic stripe MIT ID card belonging to Dennis with the track information present on Linda's card. When Dennis' card is swiped at a terminal at the Atlas Service center, Linda's Techcash account is being accessed and a picture of her displayed as shown in Figure \ref{fig:idcard}. This is a limited demonstration, however, similar attacks can be performed on credit cards. These results did not surprise us, as there are often news reports of criminals successfully skimming and cloning the magstripes of ATM cards to withdraw money under someone else’s name \cite{skimstats}. As a result, we can see that magstripe cards only afford minimal security to the cardholder - with many of the security measures being tied to physical possession of the card.

\subsection{EMV Chip Cards}

A more secure way for card-based authentication is the EMV chip, which was developed in the 2000's. EMV chip cards contain a cryptographic microprocessor that is specially designed to protect the secrets on the against unauthorized access and side channel attacks. This adds an additional layer of security through offline data authentication - a standard cryptographic check relying on public-key cryptography. The data authentication protocol is the most notable change from magstripe, as it allows the EMV chip cards to dynamically alter their responses to queries and even refuse to answer queries if they are not verified by the card issuer (ex. through offline PIN and signature verification). This provides protection against the modification or cloning of data on the EMV card. Finally, the chip itself is difficult to clone and requires expensive equipment and expertise \cite{emvhack} and physical possession of the card is often itself not sufficient for an adversary. However, the ID of the card doesn't change, so it is possible for a merchant, eavesdropper or malware in the terminal to track transactions made by the same card \cite{emvwiki}. As we do not have the resources required to mount an EMV attack, we cannot perform our own analysis of EMV security.

\section{Near-Field Communication (NFC) Payments}

In this section, we provide an overview of near-field communication (NFC) technologies and the payment systems that it enables.

\subsection{Overview of Near-Field Communication}

Near-Field Communication (NFC), much like RFID, uses electromagnetic induction between two loop antennas at a specific frequency to transmit information at a short range. The information is stored in tiny micro chips - or tags - and transmitted to readers within a certain physical range. 

Cards that make use of low frequency tags (usually transmitted at 125/134kHz) follow a security protocol similar to that of magstripe and are typically used with no additional cryptographic protection. These tags are often used when quick physical access is desired, but additional encryption is not necessary (e.g. in MIT ID cards). As such, these tags usually have a higher range (~15 inches). Some vendors tried to protect their systems by obfuscation of the design and protocol, but, in nearly all cases, this has been unsuccessful. Compared to magstripe cards, where physical contact is required to read the card, RFID cards can be read out in proximity. This leads to an additional security threat \cite{nfc-security} \cite{nfc-healthcare}. 

High frequency tags, which transmit information at 13.56 MHz and have a much tighter range of ~4 inches, can have different levels of security. If the tag only contains a UID, it has similar security to a magstripe or a low frequency tag. However, the primary advantage of high frequency tags is that they can be protected cryptographically. 

An example of a cryptographically-protected NFC card is the Mifare Classic, which is used by the MBTA for CharlieCards. In these instances, control to data sectors is protected by symmetric keys. However, the centralized nature of the data sector control means that the security of these systems can be broken if the keys are leaked. In fact, the CharlieCard systemwide keys were leaked in recent years and can now be found in a public GitHub repo. These keys can also be found by a nested attack \cite{nesting}. In the new Mifare Desfire EV1 and EMV cards, the data can be additionally protected with asymmetric keys, making them nearly uncloneable. Also, these cards are especially designed to be protected against side channel attacks, like the nested attack mentioned above. 

Recently, these NFC protocols have been applied as payment methods due to their speed, convenience, and the narrow proximity required for their use.

\subsection{NFC Card Payments}

NFC payments can be made through an NFC-enabled microchip embedded within a physical EMV chip credit card. These microchips are used to provide a fast, contactless alternative to the normal EMV chip transaction method.

While an EMV chip transaction offers additional levels of security (examples of which are discussed in Section 2.2), the additional security checks and encryption make it much slower than magstripe transactions. The NFC-enabled microchip instead allows payments to be executed and authenticated in a fraction of this time using a point-of-sale system. These systems are designed to be "secure, because [they are] only accessible in short distance" \cite{emvcontactless}. They will usually only provide the data necessary to complete the transaction with no additional capabilities (e.g. users cannot increase their card balance through the contactless interface).

EMV allows for an additional layer of authentication security by use of a PIN. However, this has slowly been phased out in the US, due to speed considerations and the growing popularity of cardless payments. In Europe, regulations mandate PIN use for transactions exceeding 20 Euros. Also, to comply with the ISO/IEC 14443 standard, RFID cards have an immutable UID, which is independent of the payment information. This enables tracking of a particular card or customer. The UID is accessible without any authentication, which can lead to privacy issues.

\subsection{NFC Cardless (Smartphone) Payments}

NFC smartphone-based payments utilize NFC chips embedded in smartphones and are becoming increasingly popular due their ease of use. Despite the many security features of NFC card-based and EMV payments, \$190 billion of transactions by over 60 million users will be made by 2020. As a result, it is imperative to understand the security implications of these systems. Mobile-based NFC payment applications such as Apple Pay and Google Pay add additional layers security to the existing NFC framework \cite{applepaynutshell}. Due the usage of the existing protocol the merchant does not require new devices. Virtually every NFC compatible payment terminal supports therefore cardless payments. These additional layers of security implementation, along with an in-depth analysis of the security policy and goals of these systems are provided in the next section. This analysis will motivate the following sections and our exploit on NFC payments.

\section{Security Analysis of Smartphone-based NFC Payments}

In this section, we perform an in-depth security analysis of smartphone-based NFC payments and describe some previous exploits on these payment technologies.

\subsection{Security Policy}

In this section, we describe a security policy for NFC cardless payment apps such as Apple Pay and Google Pay. We will define the roles and permissions of its stakeholders and describe the policies that govern data usage, protection, and sharing in order to achieve its three main goals: confidentiality, integrity, and security.

\subsubsection{Stakeholders}
Listed below are the four primary stakeholders in each NFC cardless transaction and the permissions each stakeholder has with respect to the transaction itself. We define an \textbf{adversary} in this scenario as any person outside of these four groups who wishes to listen in on, gain access to, or modify the details of the cardless transaction.

\begin{enumerate}
    \item \textbf{User:} A user is the owner of the phone and thus the NFC cardless payment method. For the purposes of this analysis, we will also assume that the user is also the owner of the physical card or payment method. The user should have the ability to perform and approve NFC-based transactions, and should also be suitably protected should they lose one of the card or phone.
    \item \textbf{Merchant:} The merchant is the receiver of the NFC cardless payment. The merchant should be able to validate the details of the transaction, but should not be able to glean sensitive information about the user's card, nor should they be able to modify the transaction details.
    \item \textbf{Device Maker (e.g. Apple):} The device maker is the creator of the phone with NFC cardless payment technology. The device maker should be able to authenticate the validity of the user's card with its issuer (see below), but should not store too much sensitive information should its security be compromised (ex. in the case of a breach that exposes all the data stored within Apple Pay).
    \item \textbf{Card Issuer:} The card issuer is the creator of the user's card. The issuer should be able to validate the user's card with the device maker (see above). Furthermore, the issuer should keep enough card information private such that the following scenario holds: suppose that the security of the user's phone $P_A$ is compromised. An adversary must not be able to take the information from phone $P_A$, transfer them to a new phone $P_B$, and be able to carry out transactions from phone $P_B$.
\end{enumerate}

\subsubsection{Security Goals}

\begin{enumerate} %enumerate
    \item \textbf{Confidentiality:} An adversary should not be able to view sensitive payment details (ex. card number), nor should they be able to track a user via a digital record of NFC cardless transactions. 
    \item \textbf{Integrity:} Transaction data should not be modified in transit between the user's phone and the merchant's payment terminal. Additionally, only a user may initiate and approve transactions, even in the case of a confidentiality breach.
    \item \textbf{Availability:} The ability to perform NFC cardless payments may not be removed by persons with fewer permissions than the user (most notably the merchant).
\end{enumerate}

\subsubsection{Security Policy}

Apple Pay and Google Pay implement several layers of security:

\begin{enumerate}
    \item \textbf{Tokenization of Card Details:} Both Apple and Google Pay do not store actual credit card data on the user's device or the cloud \cite{googlepayhowitworks}. Instead, these details are sent to the user's bank or card issuer exactly once - at the time the card is added as a payment method. The card issuer then sends back a token generated via a one-way non-reversible cryptographic hash function. This is the only number stored on the user's device and is unique to that device, providing \textbf{integrity} in the form of authentication and \textbf{confidentiality}, since neither the merchant nor an adversary can glean any details of the user's credit card number or security code off of the token. Furthermore, many devices, including iPhones, implement hardware level security for storage of these tokens, with a tamper-resistant secure element \cite{iossecurity}\cite{securelement}.
    \item \textbf{Tokenization of Transaction Details:} For each transaction, both Apple Pay and Google Pay use a one-time key to generate tokenized payment information and transmit this information to the terminal. This provides another layer of \textbf{confidentiality}, as the information used in each payment is not reusable. We also observed this phenomenon in our initial experiments - when we scanned a team member's Apple Pay card with a Proxmark3, we noticed a unique UID with each tap, even for the same virtual credit card. This was noticeably different from our experiments with an NFC contactless chip card, which had a static UID on each tap.
    \item \textbf{Additional User Authentication:} Unlike a traditional credit card, a lost or stolen iPhone is more secure. First, each transaction requires the user to authenticate and approve the payment via Touch ID. There are currently no known methods to bypass Apple's Touch ID technology to initiate a transaction from standby mode: the phone must be unlocked and authenticated via fingerprint in order to begin an Apple Pay transaction \cite{applepaywiki}. Additionally, Apple's "Find My iPhone" feature can be used to potentially recover the lost or stolen phone, and can also disable Apple Pay payment permissions on that phone should an adversary pass all other security layers. These features give the user both \textbf{integrity} (Touch ID being an additional layer of authentication) and the \textbf{accessibility} to change the permissions of that phone should it be lost or stolen.
\end{enumerate}

\begin{figure}[h]
\centering
\includegraphics[width=0.6\textwidth]{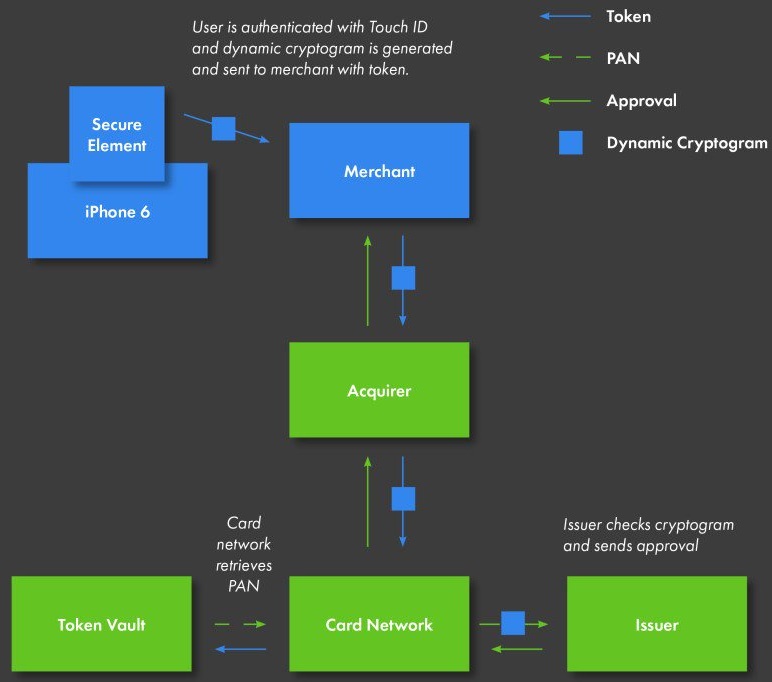}
\caption{Anatomy of an Apple Pay Transaction}
\label{fig:applepayanatomyfig}
\end{figure}

Figure \ref{fig:applepayanatomyfig} offers a simplified view of the additional measures taken "under the hood" that make a NFC cardless payment more secure than even EMV or NFC card-based payments \cite{anatomyapplepay}. 

\subsection{Previous Exploits}

As described previously, NFC itself has little inherent security. As a result, we focus this section on previous exploit attempts on smartphone-based NFC payment systems, which apply several layers of additional security. We mention that we found several attacks on apps that use Apple Pay for in-app transactions \cite{applepayreplay}, however we focus on physical transactions using NFC at a terminal.

The only attacks we found focused on using public or fake Wi-Fi hotspots. These hacks can manipulate the amount of a payment or ask a user to create a profile and steal the one-time use dynamic token. Despite Apple's assertion that these tokens can and should only be used once, some payment terminals are setup to allow the same token to be used twice \cite{wifiapplepayhack} \cite{wifiapplepayhack1}. Besides this, the most common exploit is for scammers to buy stolen consumer identities and loading them into Apple Pay to use for transactions. This enables credit card fraud, since it is more convenient then printing out a credit card to use in stores \cite{applepayscam}. Overall, there seem to be very few Apple Pay exploits that have been publicly revealed.

\section{NFC Payments Wormhole Attack}
In this section, we construct an attack, by making a 'wormhole' through which we could make a payment with the card being in a different location. In the first section, we describe two approaches to the attack, followed by limitations of wormhole attacks in the second section.
\subsection{Overview}
A scenario we imagine is that we might have a victim with an NFC enabled VISA-card or smartphone in Boston. The problem is that the card or payment information is not cloneable. Our solution is to perform an transaction using this victim's card without cloning. Our attack relies on an adversary being in near proximity to the real card. An accomplice of this adversary would be elsewhere, perhaps at a vending machine in Cambridge. The adversary in proximity and the accomplice would establish a tunnel ('wormhole') between the reader device in proximity to the victim and the card emulator that the accomplice would use to make a payment at the remote location. One might assume that such an attack might not work as the distance is too big. However, we found that no distance bounding protections are used and that the tunnel can be used to transmit data that is found by eavesdropping by the adversary in proximity. This could potentially be used to also modify the data in traffic. We describe two approaches to this attack, relying on slightly different hardware.

\subsubsection{Approach 1: Smartphone based wormhole}
Our first approach uses NFCGate, which is a NFC relay and security analysis app available for Android devices \cite{nfcgate}. Using a patched version of the \texttt{libnfc} library, which allows userspace access to NFC devices, and specific Broadcom chipsets that are present in many modern Android phones, it allows NFC cards to be read over long distance to allow for eavesdropping, protocol reverse engineering or traffic modification applications. It requires two NFC enabled Android phones and one server. One phone acts as a reader in proximity to card and the second phone acts as a card emulator. The emulator phone must support host card emulation (HCE). The server acts as a tunnel between the two phones, which are connect to the server over the internet.

\subsubsection{Approach 2: Proxmark based wormhole}
In our second approach, we replace the reader phone with a Proxmark3 \cite{proxmark}. Proxmark3 is a powerful RFID tool, that can be optimized for a range of 3-4cm and used at high and low frequencies, unlike limited range and frequency bands for phone based NFC readers. It can be used in standalone mode or connected to computer. The setup of this approach is similar to that of the approach above. In this case, the Proxmark3 is used as a reader in proximity to the real payment card and interfaces with a phone or computer to transmit data over the tunnel. However, this attack is not as discreet as the the approach above.

\subsection{Limitations of Wormhole Attacks}

Wormhole attacks have several limitations, with physical distance being an obvious one. The strength of the antenna used limits how far the reader must be from the card. For phone readers this distance is 1cm, while for the Proxmark3 this distance can reach 5cm. Because the reader is only successful in such short ranges, attackers may find it difficult to covertly read information. 

Another limitation is when cards are physically stacked on top of each other and have the same frequency tags, they interfere with each other, making it difficult for readers to accurately obtain information. ISO/IEC 14443 supports anti-collision operations making it theoretically possible to access stacked cards. But this is limited to cards with different applications, as in our case readers would try to select the payment function of the card. Also it is difficult to power on multiple cards at the same time with small, compact readers. Because many people keep their cards stacked in wallets or phone cases, it can be particularly hard to read. MIT IDs, however, have a different, lower frequency tag than most credit cards, so it can still be read through a stack of high frequency tag cards. 

Wormhole attacks are also limited by Apple Pay and Google Pay timeout features. The transaction will time out after a certain period and the user will be notified if the attacker is unable to complete payment processing within a certain time period (30 seconds for Apple Pay). This not only places time limitations on the attacker, but also brings up the problem of approval. 

Apple Pay and Google Pay both require the user to approve of a payment before authorizing the transaction, which may be done in the form of entering a password or using TouchID to unlock the phone. This means that the user is notified whenever a payment is being requested, so attackers must target users who have their phones unlocked and have already authorized a transaction, rather than just targeting users with locked phones in their pockets or out in the open.

\section{Our Wormhole Exploit}
In this section, we describe our wormhole exploit using the framework described in the first approach in section 5. We use two Android phones (an Nexus 5 and a Nexus 6P) with NFCGate. Additionally, we used a laptop to function as the NFCGate server. This laptop was connected to each of the phones via its IP address over an wireless access point we created for our experiment. However, this attack can be performed over over public Wi-Fi. We selected a vending machine that accepted contactless payment as our point of attack.

One phone was set as a card reader, and the other was set to be the card receiver. To simulate a NFC wormhole attack, we held the card receiver phone up to the vending machine's payment terminal, and held the card reader phone next to the contactless payment method but far away from the vending machine. Our goal was to authorize payment on the vending machine using the contactless payment method that was nowhere near the vending machine.

We first tried this attack with a contactless credit card, then with Android and Apple devices using their respective payment apps.

\subsection{Contactless Credit Card}
\begin{figure}[h]
\begin{center}
\includegraphics[height=0.5\linewidth]{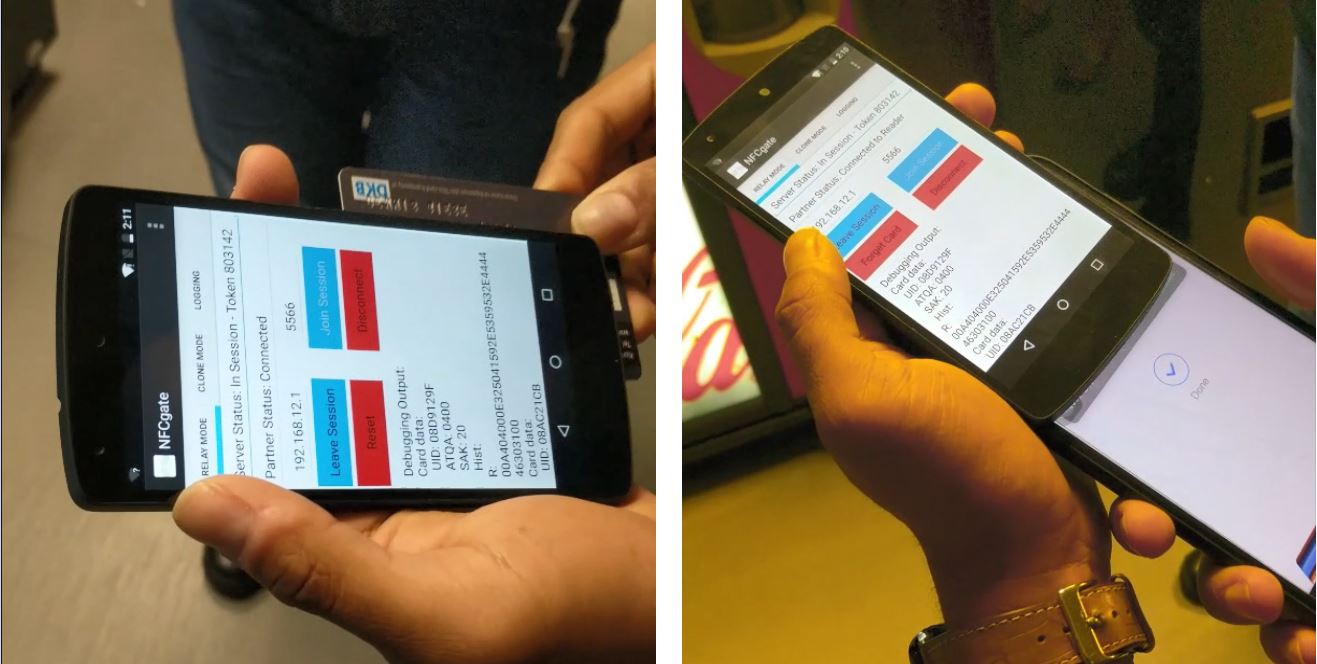}
\caption{\textit{Left}: Holding the contactless credit card to the back of the card reader phone. The phone's screen displays the captured credit card information. \textit{Right}: Holding an unlocked iPhone to the back of the card reader phone. The iPhone's screen clearly shows the ``Done'' checkmark signifying that the payment has been approved.}
\label{attacks}
\end{center}
\end{figure}

For this attack, we held the contactless credit card (VISA) to the back of the reader phone while tapping the receiver phone to the vending machine's point-of-sale. Figure \ref{attacks} shows one of the adversaries holding the credit card close to the back of the reader phone. After the normal delay, the vending machine accepted the payment and authorized a \$1.5 charge on the credit card to purchase a drink. The credit card's owner had text message alerts set up for the card, and indeed received a notification about a \$1.5 charge to the card.

In a real attack scenario, the phone would be in the adversary's pocket and they would walk close by a victim's pants pocket with the victim's wallet inside. This would allow the reader phone to activate its NFC terminal with the victim's credit cards.

\subsection{Google and Apple Pay}
We next tried the wormhole attack using Android and Apple devices, specifically targeting the digital wallet applications Google Pay and Apple Pay built into these devices. Like above, the first adversary held the unlocked victim phone to the back of the reader phone with NFCGate installed. The second adversary then held the receiver phone to the vending machine's point-of-sale. After the normal delay for contactless payments, the vending machine authorized a \$1.5 charge on the victim phone's payment app. In Figure \ref{attacks}, the checkmark and the word ``Done'' signifying that Apple Pay has allowed the payment to go through is shown, even though the phone is no where near the vending machine in the background. The same result occurred when Google Pay running on an Android device was targeted.

One added condition needed for our wormhole attack to work on the mobile payment applications was that the devices needed to be unlocked for the payment to be processed. This is an added security feature built into Google Pay and Apple Pay. In a real attack scenario, the adversary would most likely need to target and approach close to victims who are on their phones, rather than those with their phones in their pocket for this attack to work.

\section{Recommendations}
In this section, we propose several recommendations to counter the wormhole exploit we perform in this paper. Two of these recommendations are systems level recommendations to be implemented in Apple Pay, while the last recommendations are for users of these systems. A secure system would ideally implement a combination or all of these recommendations.

Our first recommendation is to implement a cryptographic distance bounding protocol, which would require the transmitting card or phone to prove that it is within some upper bound distance $d$ from the verifier, in this case the terminal \cite{distanceboundwiki}. These protocols rely on the time delay between the time challenge bits are transmitted and the time response bits are received to derive an upper bound on the distance based on speed-of-light computation. However, due to physical constraints, these can be difficult to reliably implement and might not be sufficient for wormhole attacks that are conducted nearby (e.g. between two adjacent checkout counters).

Our second recommendation for smartphone based NFC transactions is to require explicit transaction approval on phone. Rather than just requiring approval to \textit{initiate} a transaction as is currently implemented in both Android and Apple Pay, a user could be asked to accept to complete a transaction, including the transaction amount. This would require and extra push to the users phone during the transaction process and might hamper the user experience, but would provide strong protection against wormhole attacks, when a same-time wormhole attack is performed on a victim.

Our last recommendations are directed at users of these payment technologies who may be unsuspecting victims of wormhole attacks. RFID blocking wallets can prevent signals from being read as opposed to a card simply out in the open or in an unprotected wallet. Finally, simply making sure to lock phones can help to solve this problem.  

\section{Conclusion}

From our experiments, we see that a variety of magstripe and NFC security methods can all be breached, whether through card skimming or wormhole attacks. We see that magstripes are generally less secure than NFC-based payment methods, as the card information can actually be stolen, cloned, and modified using simple technologies such as a magstripe reader. NFC based systems are generally more secure, as the secure information cannot actually be stolen and read but rather only used in attacks, such as a wormhole breach.

Using two simple Android phones and a laptop functioning as a NFCGate server, we were able to perform a NFC wormhole attack using an NFC reader and tunnel an NFC transaction between a vending machine and a victim's contactless payment method with minimal difficulty and delay in transaction time. This attack was tested with both contactless credit card and the mobile payment applications Google Pay and Apple Pay.

The ease at which these attacks were carried out was eye-opening. It would not be difficult for a malicious team of attackers to download and install NFCGate on a NFC simulator with stronger signal strength than a mobile phone. This way, the adversary needs only to stand close to the victim rather than pressing the victim's payment method against the NFC simulator. This could happen in any crowded place, and with the lack of distance-bounding on any of these payment methods, the adversary's partner could be around the world paying using the victim's payment method. In addition, in US it is possible to withdraw money from an ATM using NFC, which could be another avenue in which vulnerabilities could be leveraged.

One note to make is that we noticed NFC payment methods involving mobile phones were generally more secure than contactless credit cards, as the wormhole exploit only worked when the victim's phone was unlocked and authorized to pay via Google Pay or Apple Pay. This provides an added layer of security which contactless credit cards do not have. However, with people constantly glued to their phones nowadays, this added layer of security might be meaningless in a crowded location.

\section{Future Work}
In the future, we hope to build upon our work by testing two more scenarios to make this attack more applicable to real world scenarios. 

First, we want to examine if we are able to perform a same-time-proxy payment with a wormhole attack. The scenario we imagine is that a victim is making a payment at a checkout counter. An adversary is standing next to them in the checkout and is able to read payment data from the victim and tunnel it to an accomplice at a neighboring checkout counter, who is able to complete a transaction using the victim's data at the same time. With this experiment, we would seek to observe whether both payments would go through.

Secondly, we also want to try and optimize the range from which the adversary must be from the victim. Though it is conceivable that an adversary might be right next to a victim in a checkout line, a more practical attack would allow for the wormhole attack to be perpetrated from a longer range. This limitation is mainly imposed by hardware and requires a stronger antenna. One way to overcome this limitation in a same-time attack would be to attempt to place an adversarial reader behind or in very close proximity to the terminal, so the unsuspecting victim consumer would be tapping their phone to a `trojan' reader.

These two scenarios will help determine the applicability of this wormhole attack, which we have shown is a concerning vulnerability, in more general situations. 

\bibliographystyle{ieeetr}
\bibliography{references}

\end{document}